\documentclass[aps,pra,tightenlines,floatfix,showpacs,superscriptaddress]{revtex4-1}
\usepackage{graphicx}
\usepackage{amsmath}
\usepackage{amssymb}
\usepackage{times}
\usepackage[english]{babel}
\usepackage{color}

\global\long\def\avg#1{\langle#1\rangle}

\global\long\def\dg{\dagger}

\def \< {\langle}
\def \> {\rangle}
\def \bs {\bar{\sigma}}
\def \s {\sigma}
\def \dg {\dagger}
\def \cs#1 {c_{#1 \s}}
\def \csd#1 {c_{#1 \s}^\dg}
\def \cbs#1 {c_{#1 \bs}}
\def \cbsd#1 {c_{#1 \bs}^\dg}

\begin{document}

\title{Interaction-induced conducting-nonconducting transition of ultra-cold atoms in 1D optical lattices}

\author{Chih-Chun Chien}
\email[Contact: ]{chienchihchun@gmail.com}
\affiliation{Theoretical Division, Los Alamos National Laboratory, MS B213, Los Alamos, NM 87545, USA}

\author{Daniel Gruss}
\affiliation{Department of Physics, Oregon State University, Corvallis, OR 97331, USA}

\author{Massimiliano Di Ventra}
\affiliation{Department of Physics, University of California, San Diego, CA 92093, USA}

\author{Michael Zwolak}
\email[Contact: ]{mpzwolak@gmail.com}
\affiliation{Department of Physics, Oregon State University, Corvallis, OR 97331, USA}

\date{\today}

\begin{abstract}
The study of time-dependent, many-body transport phenomena is increasingly within reach of ultra-cold atom experiments. We show that the introduction of spatially inhomogeneous interactions, e.g., generated by optically-controlled collisions, induce negative differential conductance in the transport of atoms in 1D optical lattices. Specifically, we simulate the dynamics of interacting fermionic atoms via a micro-canonical transport formalism within both mean-field and a higher-order approximation, as well as with time-dependent DMRG. For weakly repulsive interactions, a quasi steady-state atomic current develops that is similar to the situation occurring for electronic systems subject to an external voltage bias. At the mean-field level, we find that this atomic current is robust against the details of how the interaction is switched on. Further, a conducting-to-nonconducting transition exists when the interaction imbalance exceeds some threshold from both our approximate and time-dependent DMRG simulations. This transition is preceded by the atomic equivalent of negative differential conductivity observed in transport across solid-state structures.
\end{abstract}


\maketitle

\section{Introduction}
Advances in experimental studies of quantum transport of ultra-cold atoms in optical lattices \cite{Inguscio04, Porto05, Esslinger07,Blochtransport,Esslinger12} draw attention to different aspects of systems out of equilibrium. In conventional condensed matter settings, an electrical current is induced by applying a voltage difference across the sample. The interaction of the charge carriers are usually homogeneous, except near impurities. In contrast, ultra-cold atoms are charge neutral and so far the motion of an atomic cloud has been generated by placing the cloud away from its equilibrium position \cite{Inguscio04} or by a sudden shift of the minimum of its trapping potential or its distortion \cite{Porto05,Esslinger07,Blochtransport,Esslinger12}. The interactions are generated by either adding another species of atoms \cite{Inguscio04} or tuning magnetic-field controlled collisions \cite{Esslinger07,Blochtransport}.

In this paper we instead suggest a method to drive a mass current of atoms via {\it local tuning of interactions}.
This, in turn, reveals interesting phenomena otherwise difficult to observe in conventional solid state systems.
A key to realize interaction-induced transport is controllable inhomogeneous interactions -- a novel possibility offered by optically tunable collisions of ultra-cold atoms. The optical Feshbach resonance (OFR) \cite{OFR1,OFR2,OFR3,MDS} is a promising technique for controlling interactions {\it locally} using focused laser beams. For instance, the spatial modulation of density using the OFR of bosons \cite{OFR2} demonstrates the power of controllable inhomogeneous interactions. Other exotic equilibrium structures may also be generated by using this tool \cite{ChienOFR}.

We also point out that these optically controlled interactions can be \textit{time dependent}. Suddenly turning on the interactions, for instance, will drive the system out of equilibrium. Thus, instead of using external driving mechanisms such as a voltage bias, it is possible, with suitable initial conditions and patterns of interactions, to induce a mass current. We show that for reasonably weak interactions the mass current of fermions is similar to a charge current in electronic systems. Further, increasing the interaction strength leads to the atomic equivalent of negative differential conductance, where the current reaches a maximum value and then subsequently decreases. A mean-field approximation predicts that this decrease leads to a conducting-to-nonconducting transition after the interaction strength exceeds some threshold (dependent on the filling). This transition may be explained by a mismatch of energy spectra between the interacting and non-interacting parts of the system, but this argument does not rule out other possible current-carrying states.

Time-dependent density-matrix renormalization group (td-DMRG) simulations \cite{tdDMRG_RMP}, as well as a higher-order approximation, demonstrate that indeed a conducting-nonconducting transition should exist in the strongly interaction-imbalanced regime when the initial state is not far away from a band insulator. The threshold value for the transition, however, differs from different approaches. Importantly, the many-body negative differential conductance remains at all levels of approximation. This is the counterpart to negative differential conductivity observed in solid-state structures \cite{NDC1,NDC2}, where changes in the carrier density or subdivisions of the Brillouin zones cause non-monotonic dependence of the current on the external field strength. We note that, recently, negative mobility of cold atoms in temporally modulated optical lattices \cite{neg_mob} and negative differential conductance in fermion-boson mixture in optical lattices \cite{Ponomarev} have been discussed. We propose, alternatively, that one may observe negative differential conductance in a setup where the optical lattice potential is static and by tuning the interactions in real space. We focus on fermions but notice that, for the Bose-Hubbard model, Ref.~\cite{Hartmann} considers the sudden connection of a superfluid and a Mott insulator with different chemical potentials and found a mass current as well.

\begin{figure}
  \includegraphics[width=1.6in,clip]{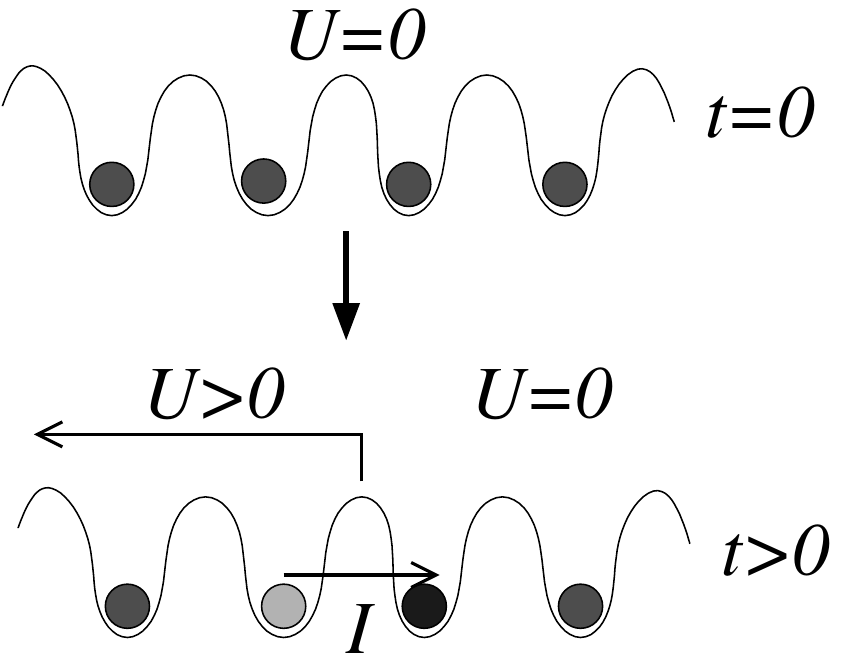}
  \caption{Schematic plot of the experimental set up we simulate: Non-interacting ultra-cold fermions are loaded into the lowest energy state of a 1D homogeneous optical lattice. At $t=0$, a focused laser beam then induces on-site repulsive interactions of the Hubbard type on the left half of the lattice. Due to the imbalance of interactions, a mass current is induced. We emphasize that there is no need to ``mechanically'' set the atoms out of equilibrium by tilting a potential nor is it necessary to introduce additional species of atoms or dissipation. Moreover, this setup may also be implemented to study quench dynamics \cite{PolkovnikovRMP}. The gray dots emphasize that atoms may be in a superposition of different quantum states.}
\label{fig:cartoon}
\end{figure}

\section{Interaction-induced transport}
In our approach -- which explicitly follows the dynamics of mass transport -- we implement the micro-canonical formalism (MCF)~\cite{Micro,Maxbook} as in Refs.~\cite{MCFshort,TDlimit} which monitors the evolution of the correlation matrix. When applied to a system of noninteracting atoms on a 1D optical lattice suddenly losing the atoms on the right half of the lattice, MCF shows very different dynamics for bosons and fermions \cite{MCFshort}. While the bosonic current decays to zero in the thermodynamic limit, the fermionic current exhibits a quasi steady-state current corresponding to a plateau in the current as a function of time. Importantly, the MCF is designed for finite closed quantum systems so it is particularly suitable for studying the dynamics of ultra-cold atoms. Although our proposed setup can be applied to bosons, to explore their dynamics theoretically one has to consider also the effect of the quasi-condensate which renders the analysis more involved. Therefore, in this paper we focus only on fermions.

We consider $N_{p\sigma}$ ($\sigma=\uparrow,\downarrow$) two-component fermions, with  $N_{p\uparrow}=N_{p\downarrow}$, loaded into a 1D optical lattice of size $N$, see Fig.~\ref{fig:cartoon}. This type of finite-length 1D lattice may be generated by inserting a thin optical barrier \cite{ring1} into a ring of optical lattice \cite{ring2}. The resulting C-shaped lattice is geometrically identical to the setup here. This confining potential creates a uniform lattice and does not require a background harmonic trapping potential parallel to the lattice. The background interactions are assumed to be negligible. The system is initially described by the tight-binding tunneling Hamiltonian $H_{0}=-\bar{t}\sum_{\langle ij\rangle,\sigma}c^{\dagger}_{i\sigma}c_{j\sigma}$, where $\langle ij\rangle$, $\bar{t}$, $c_{i\sigma}^{\dagger}$ ($c_{i\sigma}$) denote nearest-neighbor pairs, the hopping coefficient, and the creation (annihilation) operator of site $i$, respectively. The unit of time is $t_0\equiv\hbar/\bar{t}$ and is about few ms for reasonable lattice depths \cite{MCFshort}.

The initial state corresponds to the lowest energy state of $H_0$, with no correlations between up and down spin states (see, e.g., Refs.~\cite{MCFshort,TDlimit}).
The system is then set out of equilibrium by introducing interactions among atoms on the left half of the lattice by, e.g., optically-induced collisions using a focused laser beam. Here we concentrate on moderate repulsive interactions and non-integer filling. This allows us to focus on the dynamics induced by the interactions and avoid unnecessary confusions about possible equilibrium phase transitions such as the Bardeen-Cooper-Schrieffer instability or the Mott-insulating phase in  uniform and static interacting systems \cite{SmithBEC}. We model an instantaneous switch-on of the interactions with a sharp interface between the interacting and non-interacting regions and relax the former condition later on.

The Hamiltonian generating the dynamics is
\begin{eqnarray}\label{eq:He}
H_{e}&=&H_{0}+\sum_{i\in L}U\hat{n}_{i\sigma}\hat{n}_{i\bar{\sigma}}.
\end{eqnarray}
Here $\hat{n}_{i\sigma}=c_{i\sigma}^{\dagger}c_{i\sigma}$, $U$ is the onsite repulsive coupling constant, $L$ denotes the left half of the lattice, and $\bar{\sigma}$ is the opposite of $\sigma$. This model should be appropriate for moderate lattice depth \cite{Hubbard_model}. Since $\bar{t}$ can be tuned by the lattice depth and $U$ can be tuned by OFR, $U/\bar{t}$ can span a broad range so our study should be relevant to experiments.

The equations of motion for the correlation matrix are $i(\partial \langle c^{\dagger}_{i\sigma}c_{j\sigma}\rangle/\partial t)=\langle[c^{\dagger}_{i\sigma},H_{e}]c_{j\sigma} \rangle+\langle c^{\dagger}_{i\sigma}[c_{j\sigma},H_{e}]\rangle$, where $[\cdot,\cdot]$ denotes the commutator of the corresponding operators. The explicit expression can be found using standard anti-commutation relations.  After some algebra, one gets
\begin{eqnarray}
i\frac{\partial \langle c^{\dagger}_{i\sigma}c_{j\sigma}\rangle}{\partial t}&=&\bar{t}X_{\sigma}-U(\langle c^{\dagger}_{i\bar{\sigma}}c_{i\bar{\sigma}}c^{\dagger}_{i\sigma}c_{j\sigma}\rangle)_{i\in L}+U(\langle c^{\dagger}_{i\sigma}c_{j\sigma}c^{\dagger}_{j\bar{\sigma}}c_{j\bar{\sigma}}\rangle)_{j\in L} . \label{eq:fEOM}
\end{eqnarray}
Here $X_{\sigma}\equiv\langle c^{\dagger}_{i+1,\sigma}c_{j\sigma}\rangle+\langle c^{\dagger}_{i-1,\sigma}c_{j,\sigma}\rangle-\langle c^{\dagger}_{i\sigma}c_{j+1,\sigma}\rangle-\langle c^{\dagger}_{i\sigma}c_{j-1,\sigma}\rangle$ and $n_{i\sigma}=\langle \hat{n}_{i\sigma}\rangle$. We solve Eq.~\eqref{eq:fEOM} both at the mean-field level and by adding higher-order correlations.

The mean-field level solution can be obtained by Wick decomposition of $\langle c^{\dagger}_{i\bar{\sigma}}c_{i\bar{\sigma}}c^{\dagger}_{i\sigma}c_{j\sigma}\rangle$ as $\langle c^{\dagger}_{i\bar{\sigma}}c_{i\bar{\sigma}}\rangle\langle c^{\dagger}_{i\sigma}c_{j\sigma}\rangle$ since no spin-flip mechanisms are present. This is the standard Hartree-Fock approximation. The equations are closed after this approximation. Thus the dynamics after the interaction is switched on can be monitored by integrating Eq.~\eqref{eq:fEOM} with the initial condition $c_{ij\sigma}(t=0)$ given by the lowest energy state of the non-interacting Hamiltonian. We update Eq.~\eqref{eq:fEOM} in a symmetric fashion so that $\langle c^{\dagger}_{i\sigma}c_{j\sigma}\rangle=\langle c^{\dagger}_{i\bar{\sigma}}c_{j\bar{\sigma}}\rangle$ at any time. Later on we will also go beyond mean-field by developing a higher-order approximation and by performing td-DMRG simulations.

In addition to the current $I_{\sigma}=2\bar{t}\mbox{Im}\langle c^{\dagger}_{N/2,\sigma}c_{N/2+1,\sigma}\rangle$, we also evaluate the particle number on the left half lattice, $N_{L\sigma}=\langle \hat{N}_{L\sigma}\rangle$ and its fluctuations $\Delta N_{L\sigma}^{2}=\langle \hat{N}_{L\sigma}^{2}\rangle-\langle \hat{N}_{L\sigma}\rangle^{2}$, which is another property that can be determined experimentally. Here $\hat{N}_{L\sigma}=\sum_{i\in L}\hat{n}_{i\sigma}$. Explicitly,
\begin{eqnarray}
\Delta N^2_{L\sigma}&=&\sum_{i=1}^{N/2}n_{i\sigma}(1-n_{i\sigma})-2\sum_{i<j}^{N/2}|c_{ij\sigma}|^{2}.  \label{eq:c2f}
\end{eqnarray}
Right before the interactions are switched on, there can be initial number fluctuations due to the wave nature of the initial quantum state. We thus present $\Delta N_{L(\sigma)}^{2}-\Delta N_{L(\sigma)}^{2}(t=0)$ which reflects the change of the number fluctuations due to the interaction-induced dynamics.

\section{Results and discussions}
\begin{figure}
\begin{centering}
\includegraphics[width=3.4in]{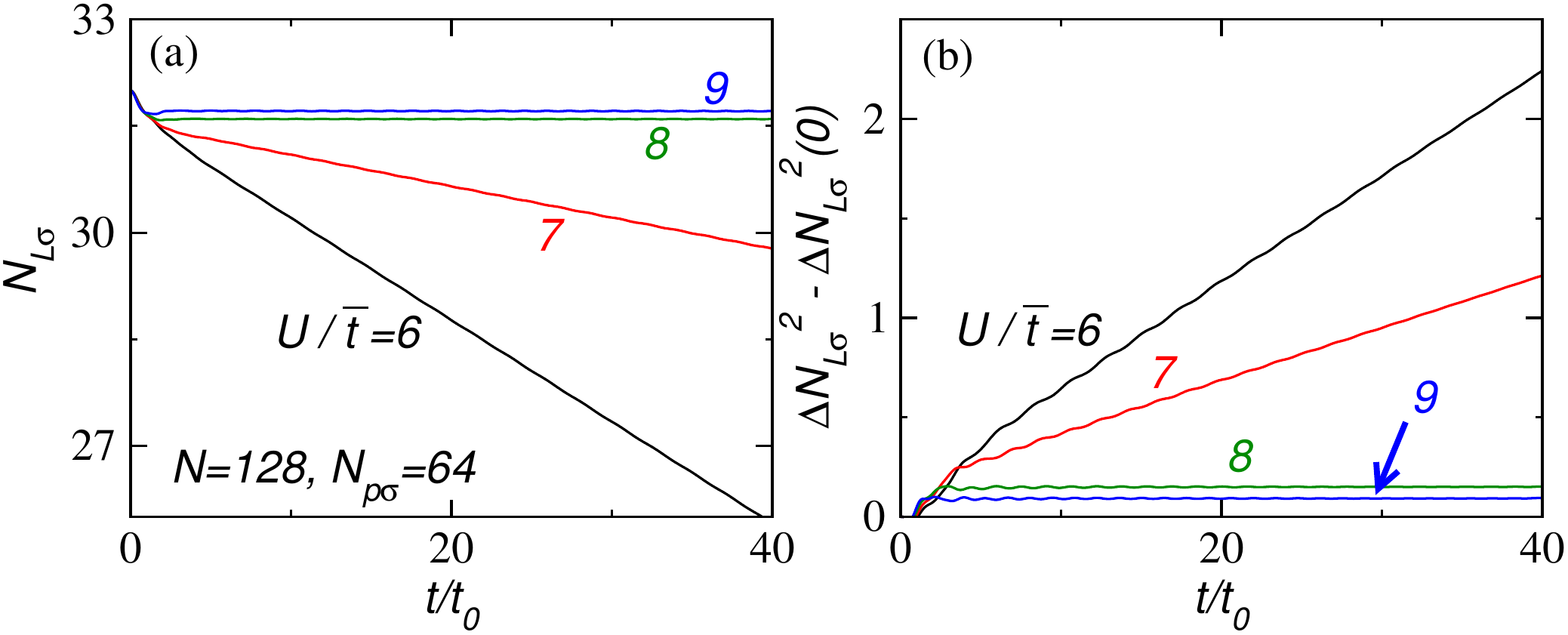}
\par\end{centering}
\caption{(Color online) (a) $N_{L\sigma}$ and (b) $\Delta N_{L\sigma}^{2}-\Delta N_{L\sigma}^{2}(t=0)$ with $N=128$ and $N_{p\sigma}=64$ for each species from mean-field dynamics.}
\label{fig:f_NdN2}
\end{figure}
Figure~\ref{fig:f_NdN2} shows the particle number and its fluctuations of the left half lattice with $N=128$ and $N_{p\sigma}=64$ for each species. Since the system size is finite, there is reflection of the current off the boundary after a revival time but we focus on dynamics before this revival occurs. As $U/\bar{t}$ increases, one can see that instead of inducing more fermions to move from the left to the right, above a certain threshold value the transport is no longer observed within a reasonable time scale (e.g., $t \le (N/2)t_0$). This corresponds to a conducting-nonconducting transition and it can be found at other ratios of $f \equiv N_{p\sigma}/N$. In general the threshold $U_c/\bar{t}$ decreases as $f$ increases. We will show that this is because for higher filling the onsite interaction energy dominates more easily.
One can see this transition even better in the average current itself.

\begin{figure}
\begin{centering}
\includegraphics[width=3.4in]{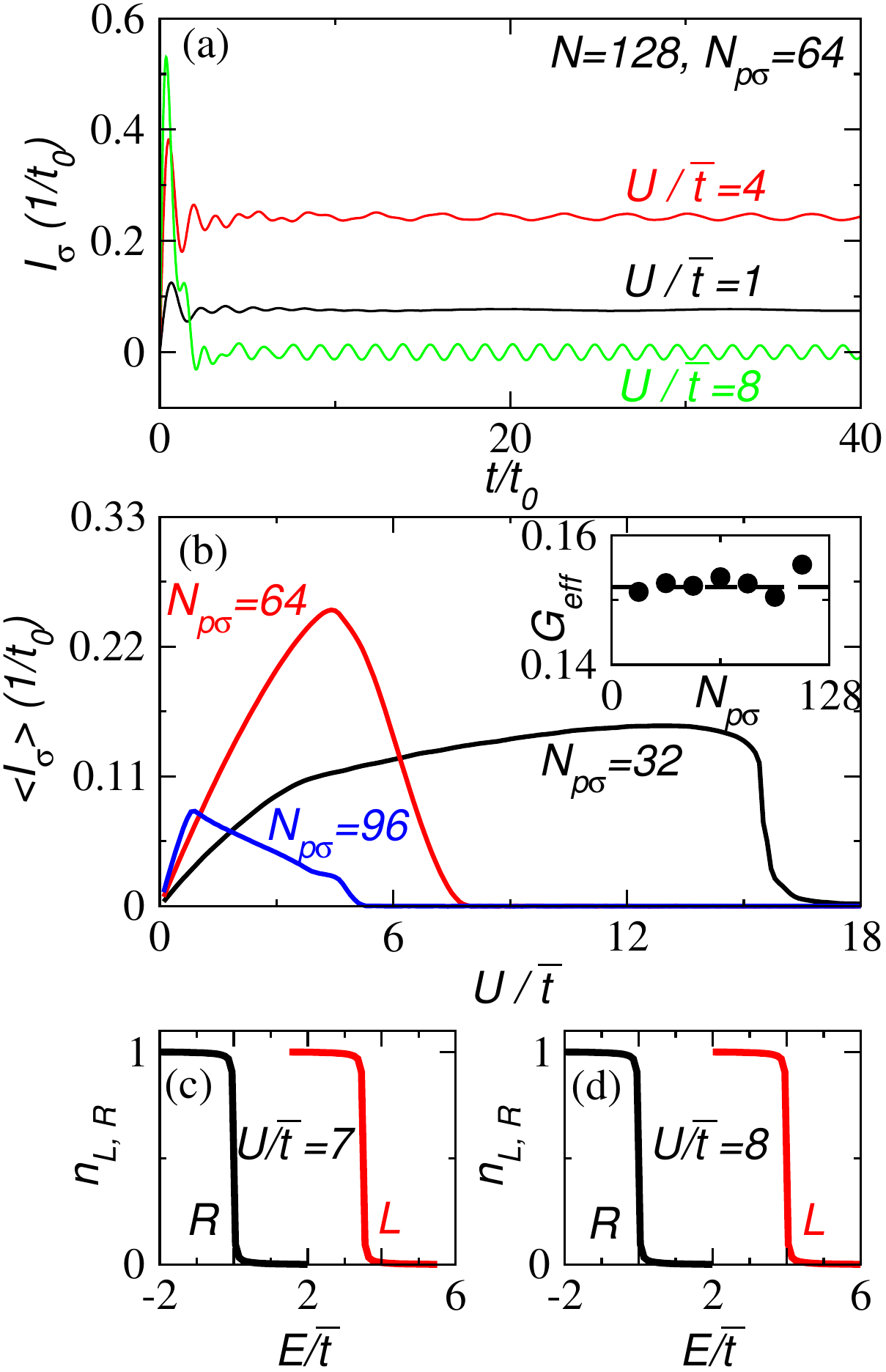}
\par\end{centering}
\caption{(Color online) (a) $I_{\sigma}$ and (b) $\langle I_{\sigma}\rangle$ of the fermionic case with $N=128$ and $N_{p\sigma}=64$ for each species from mean-field dynamics. The inset of (b) shows $G_{eff}\equiv f^{-1}d\langle I_{\sigma}\rangle/dU$ (in units of $2\pi/h$) for small $U/\bar{t}$ as a function of $N_{p\sigma}$ for $N=128$, where $f=N_{p\sigma}/N$ is the initial filling factor. The dashed line shows $G_0\approx 0.152$ in this unit. The energy spectra of the left (L) and right (R) half lattices when $U$ is switched on are shown for $U/\bar{t}=7$ (c) and $U/\bar{t}=8$ (d). Here $N=128$ and $N_{p\sigma}=64$.}
\label{fig:f_I}
\end{figure}

We note that it has been shown that with suitable initial conditions non-interacting fermions can develop a quasi steady-state current (QSSC) \cite{Bushong05,MCFshort}. This is so also for the present interacting case. The fermionic current is shown in Figure~\ref{fig:f_I}(a) for $N=128$ and $N_{p\sigma}=64$ for each species. The current may be measured by the protocol of Refs.~\cite{slowmasstran,MCFshort}.
The plateaus shown in $I_{\sigma}(t)$ indicate the existence of QSSCs. To smooth over small oscillations, we define an averaged current as $\langle I_{\sigma}\rangle\equiv (1/30t_0)\int_{10t_0}^{40t_0}dt I_{\sigma}(t)$ and plot $\langle I_{\sigma}\rangle$ as a function of $U/\bar{t}$ for $N_{p\sigma}=32, 64, 96$ in
Fig.~\ref{fig:f_I}(b). For small $U/\bar{t}$ we clearly see that the averaged current increases linearly with $U/\bar{t}$. Then the dependence deviates from a linear form before the conducting-nonconducting transition. For even larger $U/\bar{t}$, $\langle I_{\sigma}\rangle$ decreases until no finite averaged current can be found. The non-monotonic dependence of $\langle I_{\sigma}\rangle$ on $U$ resembles the negative differential conductivity observed in conventional solid-state devices \cite{NDC1,NDC2}. In the strongly interaction-imbalanced regime, we found oscillations in the density profile and further studies may connect those oscillations to the charge domains in solid-state devices exhibiting negative differential conductivity.

The interaction term thus acts as a bias on the left half lattice if $U/\bar{t}$ is small. Here we investigate the validity of this statement. By extracting the slope of the linear part of $\langle I_{\sigma}\rangle$ for small $U/\bar{t}$, one can study how efficient the interaction can induce a current. The inset of Fig.~\ref{fig:f_I}(b) shows the effective conductance $G_{eff}\equiv f^{-1}d\langle I_{\sigma}\rangle/dU$ as a function of $N_{p\sigma}$ for a fixed $N$, where $f=N_{p\sigma}/N$ is the initial filling factor. As $N_{p\sigma}\rightarrow N$ the system approaches a band insulator \cite{bi_note} and there is only a tiny range of $U/\bar{t}$ where a finite $\langle I_{\sigma}\rangle$ exists. This could contribute to some inaccuracy of the extraction of the slope close to $N_{p\sigma}\rightarrow N$. Nevertheless, as shown on the inset of Fig.~\ref{fig:f_I}(b) the effective conductance is close to the ideal quantized conductance $G_0=(2\pi \hbar)^{-1}$ for arbitrary initial filling. This establishes the feasibility of using a weak inhomogeneous interaction as an efficient driving force for inducing a current and its correspondence with a bias, $\Delta\mu=f U$.

\begin{figure}
\begin{centering}
\includegraphics[width=3.4in]{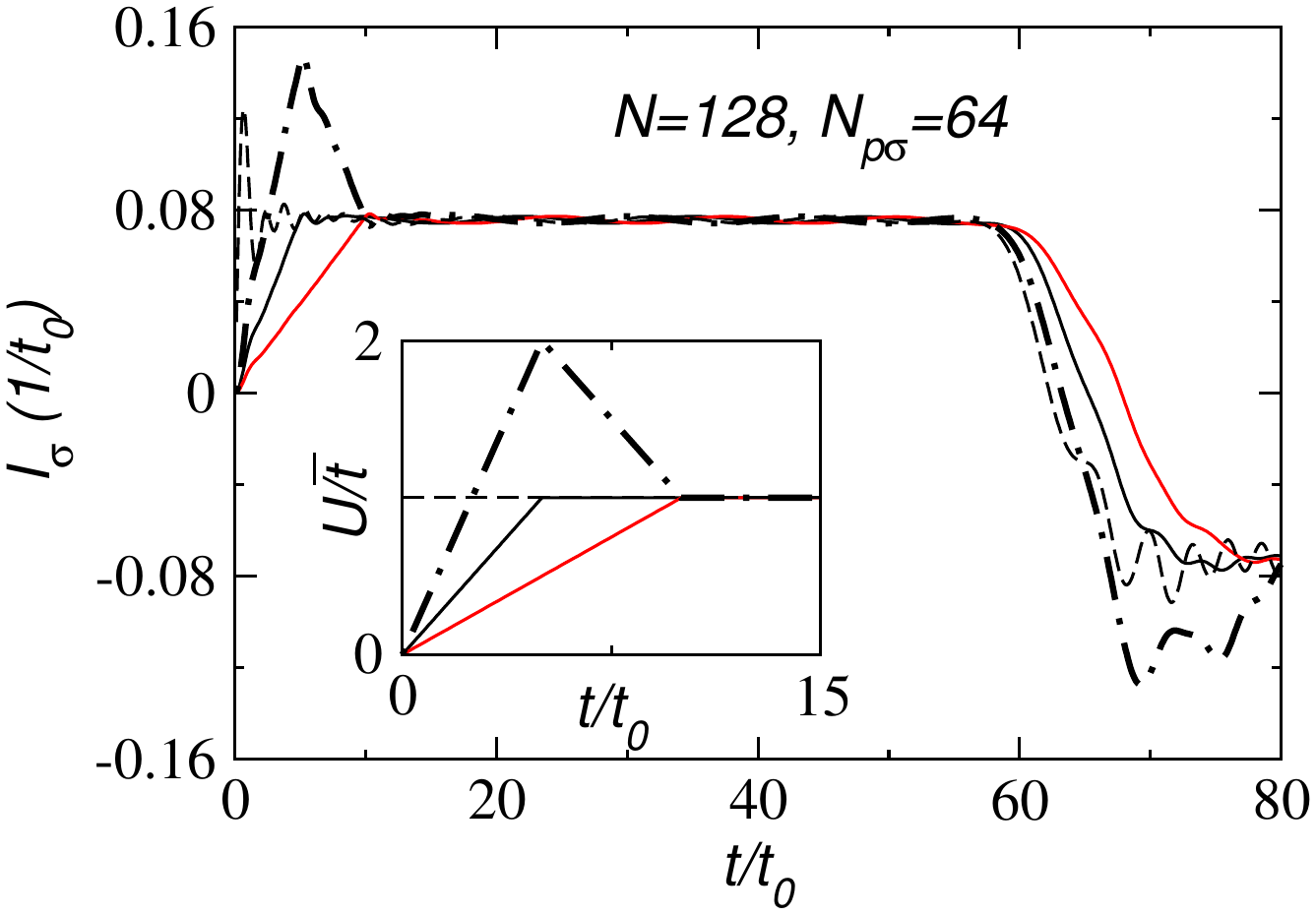}
\par\end{centering}
\caption{(Color online) Currents induced by different ways of switching-on the interaction from mean-field dynamics: A sudden switch-on of the interaction (dashed line), linear switch-ons with switching times $t_m=5t_0$ (black) and $t_m=10t_0$ (red), and a multi-step switch-on (dot-dash line). The inset shows the time dependence of $U$ for the four cases. Here $N=128$ and $N_{p\sigma}=64$.}
\label{fig:mem}
\end{figure}
In the weak interaction regime, we found another interesting phenomenon. By considering different ways of switching on the interaction, one may wonder whether the system will reach the same magnitude of the QSSC. We simulate different switching-on scenarios by introducing a time scale $t_m$ such that the interaction grows from zero to its full value during $t_m$ and remains a constant after that. Figure~\ref{fig:mem} shows that despite different time-dependences of the interaction, the system always reaches the same magnitude of the QSSC. We have also checked other more complicated functional forms for switching on the interaction and found the same QSSC. Importantly, even if the system is over-excited by a spike in the time dependence of $U$, when $U$ comes back to a constant value the height of the QSSC coincides with the case of a sudden switch-on of the interaction as shown on Fig.~\ref{fig:mem}. Thus the magnitude of the QSSC is robust (memory-less) against different ways the interaction is switched on. This feature makes the interaction-induced transport appealing for making devices since a steady output (the QSSC) is insensitive to the transient behavior of the driving force (the interaction).

When the optically-induced interactions are switched on, it is possible that the onsite potential could be shifted due to the increase of the kinetic energy of the atoms interacting with the incoming photons. We study possible effects by adding an extra term $V\sum_{i\in L,\sigma}\hat{n}_{i\sigma}$ to $H_e$ of Eq.~\eqref{eq:He} to simulate this effect with a positive $V$. This shift of the onsite potential acts like a bias so one might expect that more current will flow to the other side. However, the negative differential conductance and the mean-field conducting-nonconducting transition are robust against this positive potential shift. The energetic mechanism described below will make it clear why a positive potential shift will not affect the results qualitatively, thus showing that it is not sensitive to this type of modification of the Hamiltonian.

We now discuss the reason behind the negative differential conductance and the conducting-nonconducting transition, which is energetic in nature. In order to obtain a ``macroscopic'' steady-state current, $\alpha N$
particles must be transferred from the left to the right halves, where $\alpha$ is some small proportionality constant. However, this entails a change in energy, as the particle density decreases on the site with interactions, and this effect must be compensated by an energy change due to rearrangement of the single particle states. To be more concrete, immediately after interactions are turned on, the system is still in the ground state of $H_0$. However, the energy has shifted to
\begin{equation}
E=E_{0}+U\sum_{i\in L}\avg{n_{i\sigma}n_{i\bar{\sigma}}}_{0}=E_{0}+U\sum_{i\in L}\avg{n_{i\sigma}}_{0}\avg{n_{i\bar{\sigma}}}_{0},
\end{equation}
where Wick's theorem has been applied since the system is in the ground state of $H_0$. The subscript of $\langle\cdots\rangle$ denotes the time when the expectation is taken. For half-filling, this gives an energy $E=E_0+UN/8$. For arbitrary filling $f$, it is
\begin{equation}
E \approx E_0+f^2 U N/2, \label{eq:Ebefore}
\end{equation}
where the expression is approximate since not all site occupation numbers will be exactly $f$ (for each spin component) .

The energy at some later time will still be $E$ (since no further changes have been made to the Hamiltonian), but will have the form
\begin{equation}
E=\avg{H_0}_t+U\sum_{i\in L}\avg{n_{i\sigma}}_{t}\avg{n_{i\bar{\sigma}}}_{t} \label{eq:Etdecomp}
\end{equation}
in the mean-field approximation. Assuming the $\alpha N$ atoms are taken uniformly from the left half, the energy is
\begin{equation}
E \approx \avg{H_0}_t+f^2 U N/2- \alpha f U N. \label{eq:Eafter}
\end{equation}
Taking the difference of Eq.~\eqref{eq:Eafter} and Eq.~\eqref{eq:Ebefore}, and using energy conservation, gives $\avg{H_0}_t-E_0 \approx \alpha f U N$.
However, the quantity $\avg{H_0}_t-E_0$ can not be arbitrary, since we have a finite bandwidth. Regardless of whether one has an interacting or noninteracting state, the largest energy change due to the $H_0$ contribution is $4 \bar{t} \alpha N$, corresponding to taking the $\alpha N$ particles from the lowest part of the band (on the left half) to the highest part (on the right half). This gives
\begin{equation}
\alpha f U N \lesssim 4 \bar{t} \alpha N \longrightarrow U \lesssim 4 \bar{t}/f
\label{eq:CU}
\end{equation}
and thus results in a threshold value of $U$. Beyond this value of $U$, a macroscopic current can not flow within the mean-field approximation and when the particles are taken uniformly from the interacting lattice. We remark that here we consider the lowest energy band of the optical lattice. If the interaction energy exceeds the band gap separating different bands, the system may re-enter a conducting state.
We also remark that energy mismatches can results in other interesting phenomena as discussed in Refs.~\cite{Branschadel,DeMarco2BEC}.

To demonstrate this energetic mechanism, Fig.~\ref{fig:f_I}(c) and (d) show the energy spectra of the left (L) and right (R) half lattices when $U$ is switched on for $N=128$ with $N_{p\sigma}=64$. When $U/\bar{t}=7$ (panel (c)), there is still an overlap between the two spectra and exchanging atoms in an energy-conserving fashion is possible. For $U/\bar{t}=8$ (panel (d)), there is no overlap between the two spectra and the system evolves into a nonconducting state. This also explains why the phenomenon of negative differential conductance is robust against an additional positive onsite bias. Any further increase of the onsite energy separates the two energy spectra of the left and the right sides even farther so the nonconducting state remains. Importantly, this energetic mechanism applies to other types of setups for studying transport phenomena. For example, by introducing a step-function bias to the lattice potential and tuning the bias, there is also a conducting-nonconducting transition of the same nature (see Appendix \ref{app:biasvsint}).

However, other possibilities may occur in interacting systems for large $U/\bar{t}$ when a homogeneous conducting state is no longer favorable. From Eq.~\eqref{eq:Etdecomp} and the discussion below Eq. \eqref{eq:CU}, states could be generated that have an inhomogenous density which would give enough ``energetic relief'' to allow a macroscopic current to flow \cite{density_note}. In the mean-field solution, however, such current-carrying state could not be found. Here we emphasize that the nonconducting state is \textit{dynamically generated} and thus is very different from the widely-discussed Mott insulating phase at integer fillings, which is an equilibrium ground state. 
We now consider a higher-order approximation and also td-DMRG simulations.

\section{Higher-order correlations and td-DMRG}\label{app:higher_order}

In order to determine the accuracy of the mean-field simulations above, we also perform simulations with higher-order correlations and with td-DMRG \cite{tdDMRG_RMP}. In the higher-order approximation, we keep two-particle correlations, rather than truncating at the single-particle level. This procedure is described in Appendix \ref{app:higher}. The details of the td-DMRG simulations are given in their respective figure captions. We note that td-DMRG has been applied to study the dynamics of different cold-atom systems \cite{Fabian09}.

\begin{figure}
  \includegraphics[width=3.5in,clip]{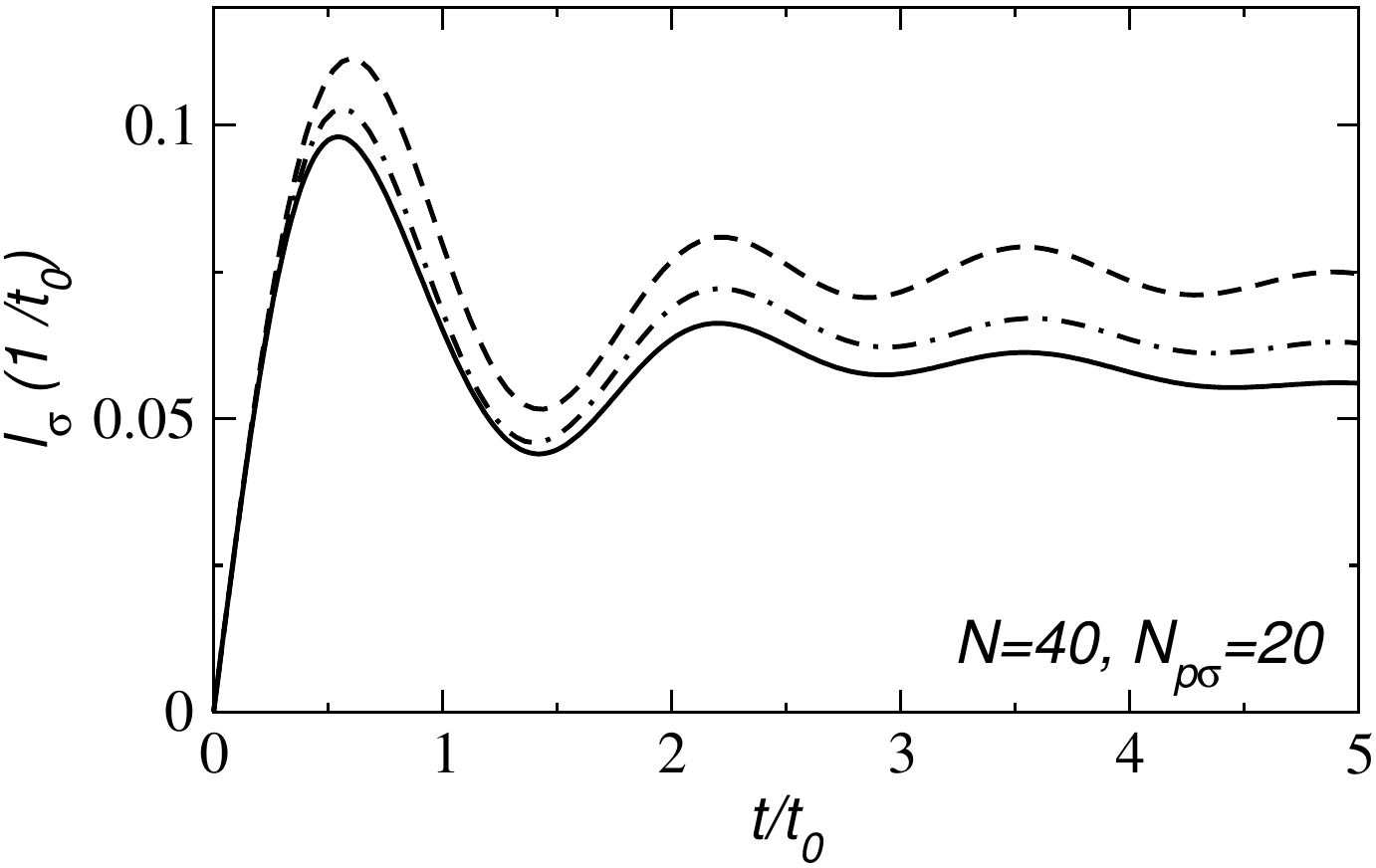}
  \caption{(Color online) Current versus time for the mean-field approximation (dashed line), the higher-order approximation including two-particle correlations (dash-dot), and td-DMRG (solid line) for $U/\bar{t}=2$. Here $N=40$ and $N_{p\sigma}=20$. For td-DMRG, we did a series of simulations using matrix-product state (MPS) dimensions 500, 2000, and 3000. The td-DMRG results shown are converged.}
\label{fig:plotNJP1}
\end{figure}

Figure~\ref{fig:plotNJP1} shows the results for $U/\bar{t}=2$, $N=40$, and $N_{p\sigma}=20$ for the three different methods. They all agree qualitatively, giving rise to a quasi steady state. Keeping track of the higher-order correlations significantly improves the accuracy of the simulation. Furthermore, the results suggest that for small $U/\bar{t}$ the transient time to reach a quasi steady state is only a few $t_0$ while the duration of the QSSC can extend much longer. Thus for reasonably large lattices, the observation of an interaction-induced QSSC and the conducting-nonconducting transition is feasible.

\begin{figure}
  \includegraphics[width=5.5in,clip]{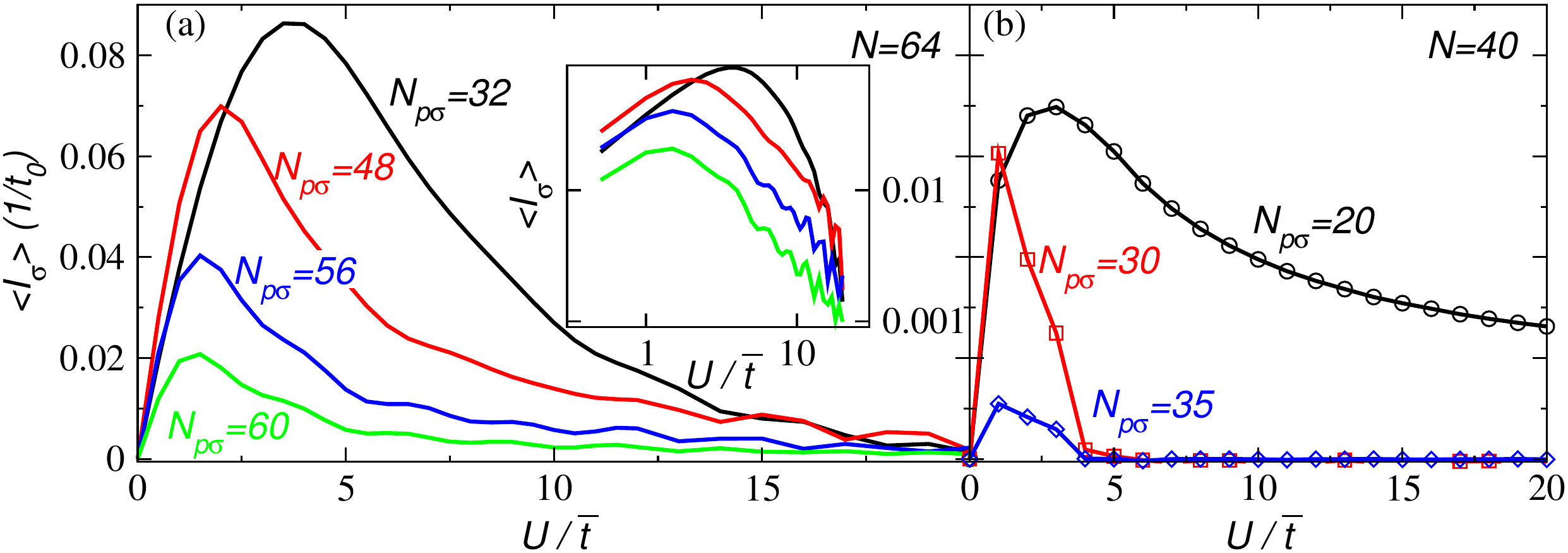}
  \caption{(Color online) Average current versus $U$. (a) $\langle I_{\sigma}\rangle$ of the fermionic case with $N=64$ and $N_{p\sigma}=32,48,56,60$ (labeled next to each curve) from the simulations with two-particle correlations. The averaged current is computed from $t=0$ to $t=2.5 t_0$. The inset of the plot shows the average current plotted on a log scale, which shows that it does indeed vary as $1/U^\alpha$ where $\alpha$ tends towards 1 as the filling increases ($\alpha\approx 2.95,1.59,1.43,1.29$ for $N_{p\sigma}=32,48,56,60$ respectively). (b) $\langle I_{\sigma}\rangle$ of the fermionic case with $N=40$ and $N_{p\sigma}=15,20,30,35$ (labeled next to each curve) from td-DMRG with MPS dimension 500. Simulations with higher MPS dimension indicate that these are accurate to within about $5\% $. The averaged current is computed from $t=5.0t_0$ to $t=10.0 t_0$. Similar to Figure~\ref{fig:f_I}, the average current initially increases with $U$ and then decreases, thus giving negative differential conductance. The td-DMRG simulations indicate that a conducting-nonconducting transition can exist for large $U$ for fillings larger than half filling.}
\label{fig:plotNJP3}
\end{figure}

Figure~\ref{fig:plotNJP3} shows the average current versus $U$. We note that inclusion of higher order correlations limits our analysis to shorter times than above. This means that the averaging partially truncates oscillations and the effect of this partial truncation is to give a finite value to $\< I_\sigma \> $ even when a longer average would give zero. This will be more prominent for shorter averages, i.e., in the absence of transient effects, one expects that the truncation will give a residual contribution of  $\< I_\sigma \> \propto 1/U$, as $U$ gives the frequency scale of oscillations (for large $U$). The inset of Figure~\ref{fig:plotNJP3} gives this tail: for filling greater than $1/2$, the average current is consistent with this interpretation. For filling equal to $1/2$, the results are ambiguous.  Most importantly, all methods clearly display the negative differential conductance, and, thus, this phenomenon should be observable in experiments. 

If the initial state is not far from a band-insulator (see the cases with $N=40$ and $N_{p\sigma}=30, 35$), the td-DMRG simulations show a conducting-nonconducting transition for large $U$. Thus this dynamically generated conducting-nonconducting transition should be observable in experiments. For the half-filling initial state, a current carrying state is observed within our simulation range, but the magnitude of the current in this state is decreasing as $U$ is increased as happened with the higher order correlation case.

\section{Conclusion}
In summary, we have shown that one may use time-dependent inhomogeneous interactions of ultra-cold atoms to explore non-equilibrium physics not easily realizable in conventional solid-state setups. For weak-interactions, the response of the system is similar to condensed matter systems where a bias is applied. This response is robust against possible transient behavior of the driving force. The application of inhomogeneous interactions gives rise to negative differential conductance, which is a many-body, atomic analog of this phenomenon in solid-state systems. Furthermore, a dynamically generated conducting-nonconducting transition is predicted from different simulations and its observation in experiments could provide another example of nonequilibrium phase transitions. Our work sheds light on the physics of complex systems out of equilibrium and may help in the design of devices in the thriving field of atomtronics \cite{atomtronics}.

CCC acknowledges the support of the U. S. DOE through the LANL/LDRD Program.
MD acknowledges support from the DOE grant DE-FG02-05ER46204 and UC Laboratories.

\appendix
\section{Comparison of bias vs. interaction}\label{app:biasvsint}
\begin{figure}
  \includegraphics[width=2.5in,clip]{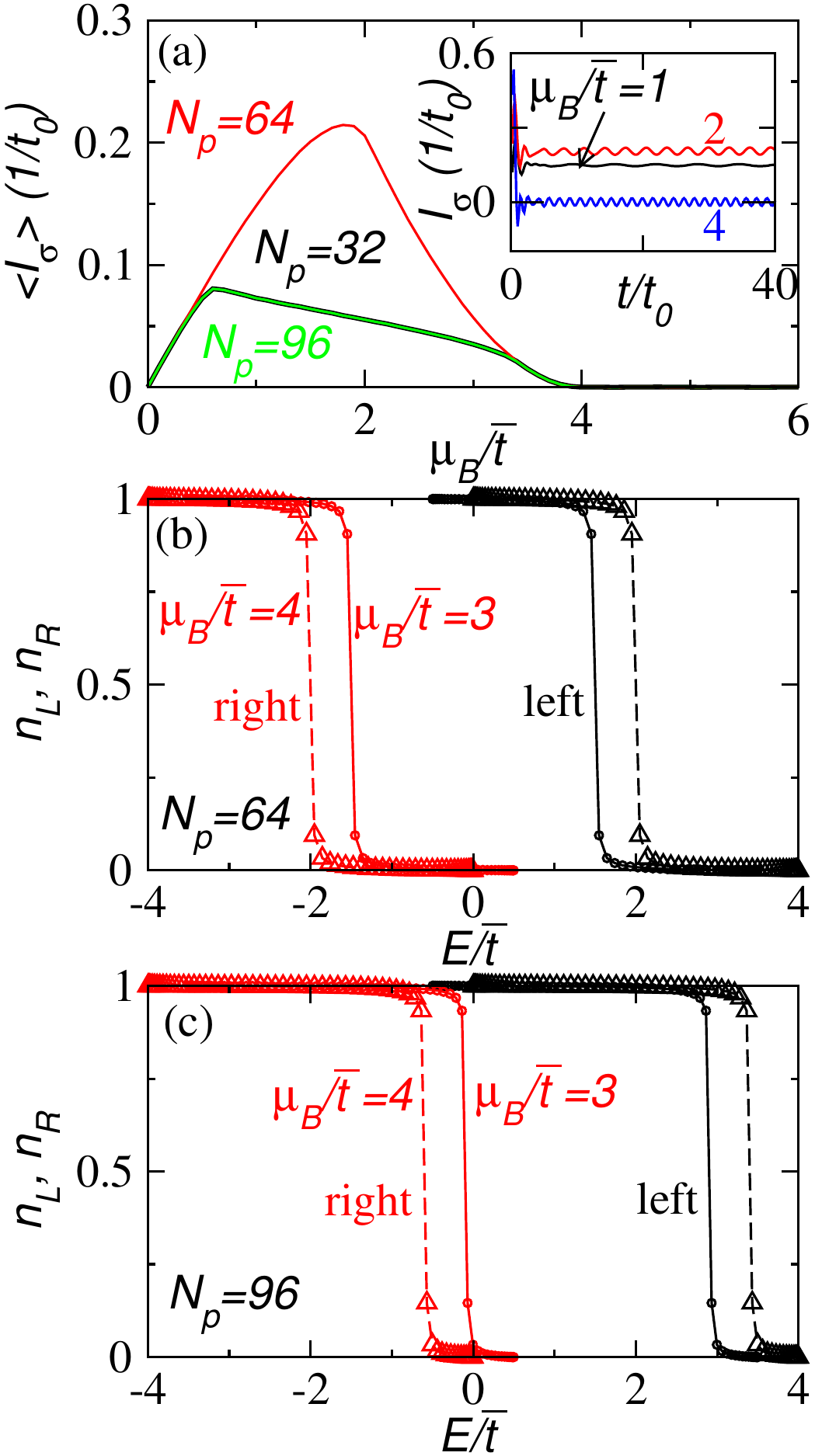}
  \caption{(a) $\langle I_{\sigma}\rangle$ as a function of $\mu_B$ for $N=128$ and $N_p=32,64,96$. The curves for $N_p=32, 96$ coincide due to the particle-hole symmetry. Inset: The currents for $\mu_{B}/\bar{t}=1,2,4$ (labeled next to each curve) with $N_p=64$. (b) Particle distributions of the left half (black) and the right half (red) for $\mu_B/\bar{t}=3$ (solid line) and $4$ (dashed lines). The symbols show energy levels. Here $N=128$ and $N_p=64$. (c) Same as (b) but with $N_p=96$.}
\label{fig:I_bias}
\end{figure}
When a bias is applied to a single-species noninteracting fermions in a uniform 1D lattice, the Hamiltonian is $H_{B}=-\bar{t}\sum_{\langle ij\rangle}c^{\dagger}_{i}c_{j}+\frac{\mu_B}{2}\sum_{i\in L}c^{\dagger}_{i}c_{i}-\frac{\mu_B}{2}\sum_{i\in R}c^{\dagger}_{i}c_{i}$. We assume the initial state is the ground state of the Hamiltonian with zero bias ($\mu_B=0$) and then it evolves according to $H_B$. For $N=128$, we define $\langle I_{\sigma} \rangle=(1/30t_0)\int_{10t_0}^{40t_0}dt I_{\sigma}(t)$, where $t_0\equiv \hbar/\bar{t}$. Figure~\ref{fig:I_bias} (a) shows $\langle I_{\sigma} \rangle$ as a function of $\mu_{B}$. One can see that above $\mu_B/\bar{t}=4$ no finite averaged current can be found. There is a conducting-nonconducting transition and the upper limit is independent of $N_p/N$, which can be seen by a similar analysis to Eqs. (4)-(8) in the main text.

As with the interaction case, energy conservation prohibits particles on the left with energy higher than the highest energy level of the right from entering the right half of the lattice. This occurs when $\mu_B$ is larger than the bandwidth of the original uniform lattice, which is $4\bar{t}$. The particle distributions of the two half lattices when the interaction is switched on can be found as follows. The initial correlation matrix $c_{ij}(t=0)$ of the whole lattice is $\sum_{p=1}^{N_p}U_0^{\dagger}(i,p)U_0(p,j)$, where $U_0$ is the unitary matrix that diagonalizes the hopping Hamiltonian. After finding $U_L$ and $U_R$ that diagonalize the Hamiltonian of the left and the right half lattices with eigenvalues $E_{L}$ and $E_{R}$, the particle distributions are given by $n_{L/R}(k)=\sum_{i,j \in L/R}(U_{L/R}^{\dagger})_{kj}U_{ik}c_{ij}(t=0)$, where $k=1,\cdots,N/2$ labels the energy levels $E_{L/R}(k)$. Fig.~\ref{fig:I_bias} (b) and (c) show the particle distributions for the left and right half lattices for $\mu_B/\bar{t}=3$ and $4$ with $N_p=64$ and $96$. One can see that when the bottom of the left half energy band moves above the top of the band of the right, no current is expected since energy is conserved.

\begin{figure}
  \includegraphics[width=2.5in,clip]{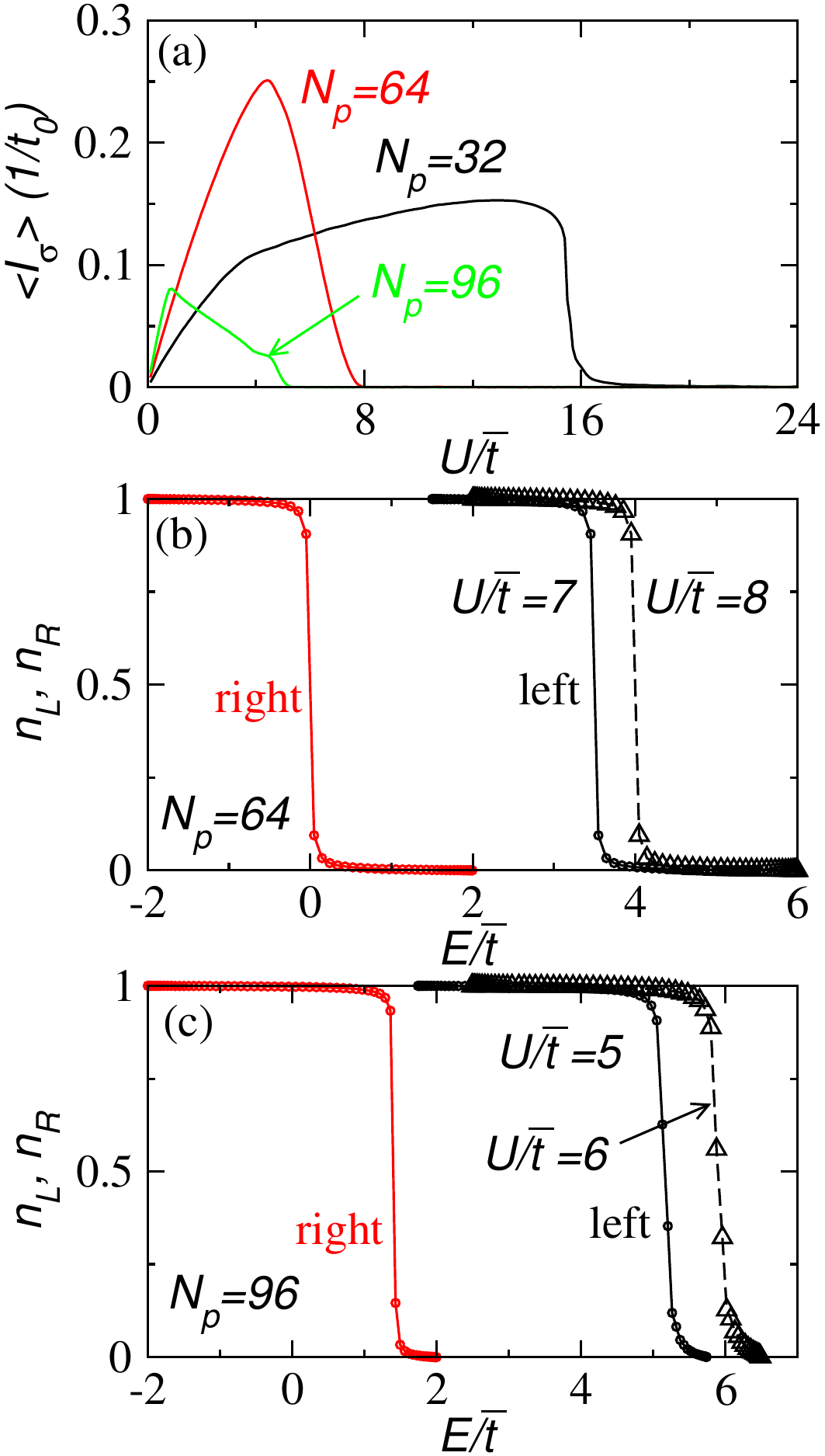}
  \caption{(a) $\langle I\rangle$ as a function of $U$ for $N=128$ and $N_p=32,64,96$ from the mean-field approximation. (b) Particle distributions of the left half (black) and the right half (red) for $U/\bar{t}=7$ (solid line) and $8$ (dashed line). The symbols show energy levels. Here $N=128$ and $N_{p\sigma}=64$. (c) Same as (b) but $N_{p\sigma}=96$ and $U/\bar{t}=5$ (solid line) and $6$ (dashed line).}
\label{fig:halfU_dist}
\end{figure}
Taking the same definition for $\langle I_{\sigma} \rangle$ for the interaction-induced case, we plot $\langle I_{\sigma} \rangle$ as a function of $U$ in Fig.~\ref{fig:halfU_dist}. At the mean-field level there is also a conducting-nonconducting transition, but the threshold value depends on the filling $N_{p\sigma}/N$. The left half lattice has Hamiltonian $H_L=\sum_{\langle ij\rangle \in L,\sigma}c^{\dagger}_{i\sigma}c_{j\sigma}+U\sum_{i\in L}\hat{n}_{i\sigma}\hat{n}_{i\bar{\sigma}}$. Here $\hat{n}_{i}=c^{\dagger}_{i\sigma}c_{i\sigma}$. To find the energy levels of $H_L$ at $t=0$, we replace $\hat{n}_{i\bar{\sigma}}$ by its expectation value, which is known from $c_{ij,\sigma}(t=0)$ and we assume the particle number of the two species are equal. After diagonalizing $H_L^{eff}=\sum_{\langle ij\rangle \in L,\sigma}c^{\dagger}_{i\sigma}c_{j\sigma}+U\sum_{i\in L}n_{i\bar{\sigma}}\hat{n}_{i\sigma}$, we found $E_L(k)$ with $k=1,\cdots,k$ and the unitary matrix $U_L$. The right half lattice remains noninteracting and one can find $E_R(k)$ and $U_R$. The distribution functions are then given by $n_{L/R}(k)=\sum_{i,j \in L/R}(U_{L/R}^{\dagger})_{kj}U_{ik}c_{ij}(t=0)$.

Fig.~\ref{fig:halfU_dist} (b) and (c) show the distributions of the right and left half lattices for $N_{p\sigma}=64$ and $96$ and $N=128$. The nonconducting states emerge slightly above $U/\bar{t}=7$ ($U/\bar{t}=5$) for $N_{p\sigma}=64$ ($N_{p\sigma}=96$). The distributions for $U/\bar{t}$ below the threshold value still show observable overlaps between the filled states of the left half lattice and the empty states of the right half lattice. When $U/\bar{t}$ is above the threshold value, the distributions of the two half lattices are separated so energy conservation forbids transport at the mean-field level. We thus observe a nonconducting state due to energy-level mismatches of the two half lattices in bias-induced transport, where they are exact solutions, and in interaction-induced transport at the mean-field level.

\section{Higher-order approximation}\label{app:higher}

In this approximation, we preserve the two particle correlation functions, i.e., we do not use a Wick decomposition at the two-particle level. Rather, we decompose the three-particle correlation functions into products of single-particle correlation functions and two-particle correlation functions, e.g.,
\begin{eqnarray}\label{eq:4th}
\< \csd{i} \cs{j} \csd{k} \cs{l} \cbsd{m} \cbs{n} \> &= &\frac{1}{3} \left(
\< \cbsd{m} \cbs{n} \> \< \csd{i} \cs{j} \csd{k} \cs{l} \> \right. \nonumber \\ & &\left.
+ \< \csd{i} \cs{j} \> \< \cbsd{m} \cbs{n} \csd{k} \cs{l} \> - \< \csd{i} \cs{l} \> \<    \csd{k} \cs{j} \cbsd{m} \cbs{n} \> + \delta_{jk} \< \csd{i} \cs{l} \> \< \cbsd{m} \cbs{n} \> \right. \nonumber \\ & &\left.
- \< \csd{k} \cs{j} \> \< \csd{i} \cs{l} \cbsd{m} \cbs{n} \> + \< \csd{k} \cs{l} \> \< \csd{i} \cs{j} \cbsd{m} \cbs{n} \> + \delta_{jk} \< \csd{i} \cs{l} \cbsd{m} \cbs{n} \> \right) .
\end{eqnarray}
This allows us to include all of the two-particle correlation functions in the simulation. Since there are three ways to decompose the three-particle correlation functions, we chose to average the contributions of each in order to increase the numerical stability of the simulation. We note that inclusion of the two-particle correlations is only a short time approximation and the simulations can become unstable for intermediate to long times. Thus we have restricted these simulations to short times. The results for the average current presented in the main text are there therefore limited to relatively short time scales for large values of $U$. This means that the averaging partially truncates oscillations and there will be some artifacts associated with this truncation.

\bibliographystyle{apsrev4-1}

\begin{thebibliography}{33}
\bibitem [{bi_()}]{bi_note}%
  \BibitemOpen
  \href@noop {} {}\bibinfo {note} {For a single-band model of fermions
  considered here, $N_{p\sigma}/N=1$ corresponds to a band insulator and no
  current can flow even in the presence of inhomogeneous
  interactions.}\BibitemShut {Stop}%
\end{thebibliography}
%

\end{document}